\documentclass[useAMS,usenatbib,usedcolumn,usegraphicx]{mn2e}

\def\etal{{\it et\thinspace al.}\ }
\def\p3{{[{\sc P\,iii}]}}

\def\mum{{$\mu$m}\ }

\def\eion{{(e~+~ion)}\ }

\def\ne{{N$_e$}}

\usepackage[dvips]{graphics}
\usepackage[dvipsnames,usenames]{color}

\newcommand{\be}{\begin{equation}}
\newcommand{\ee}{\end{equation}}

\title{Collision strengths for  FIR and UV transtions in
{\sc P\,iii} and the phosphorus abundance}
\author
[Rahla Naghma, Sultana N. Nahar, Anil K.\ Pradhan]
       {Rahla Naghma$^1$, Sultana N. Nahar$^1$, Anil K.
Pradhan$^{1,2,3}$\\
       $^1$ Department of Astronomy, $^2$ Chemical Physics Program, $^3$
Biophysics Graduate Program,
 The Ohio State University, Columbus, OH 43210, USA.}
\date{Accepted  xxxxxx 
      Received xxxxxx;
      in original form xxxxxx}

\pagerange{\pageref{firstpage}--\pageref{lastpage}}
\pubyear{2011}

\def\LaTeX{L\kern-.36em\raise.3ex\hbox{a}\kern-.15em
    T\kern-.1667em\lower.7ex\hbox{E}\kern-.125emX}

\begin{document}

\maketitle

\label{firstpage}

\begin{abstract}
 Phosphorus abundance is crucial for DNA-based extraterrestrial life in
exoplanets. Atomic data for observed spectral lines of P-ions are
needed for its accurate determination. 
We present the first calculations for collision strengths
for the forbidden \p3 fine structure transition $3s^23p (^2P^o_{1/2-3/2})$
within the ground state at 17.9 \mum, as well as allowed UV
transitions in the $3s^23p (^2P^o_{1/2,3/2}) \rightarrow 3s3p^2
(^2D_{3/2,5/2}, ^2S_{1/2}, ^2P_{1/2,3/2})$
multiplets between 915-1345 $\AA$. Collision strengths are computed using
the Breit-Pauli R-Matrix method including the first 18 levels, and they
exhibit extensive auto-ionizing resonance structures. In particular, the
Maxwellian averaged effective collision strength for the FIR 
17.9 \mum transition 
shows a factor 3 temperature variation broadly peaking at typical nebular
temperatures. Its theoretical emissivity 
with solar phosphorus abundance is computed relative to
H$\beta$ and found to be similar to observed intensties from
planetary nebulae; the abundances derived in earlier works
are 3-5 times sub-solar. The results pertain to the reported
paucity of phosphorus from preferred production sites in supernovae,
and abundances in planetary nebulae and supernova remnants.

\end{abstract}


\section{Introduction}
 
 Phosporus abundance is of considerable interest in the search for life
forms in exoplanets. It is the backbone element in the DNA molecule,
enabling chemical bonds among myriad nucleotides that constitute the
complex double-helical structure. However, ascertaining
nucleosynthesis pathways and determining the 
actual abundance of phosphorus (Z = 15) is challenging because it is much lower
than low-Z $\alpha$-elements, and orders of magnitude lower
than the other five most common elements of the six 
that constitute the CHONPS-based life organisms on the Earth. 
The photospheric 
solar abundances numerically relative to hydrogen are
(Asplund \etal 2009): C (2.7 $\times 10^{-4}$), N (6.8 $\times 10^{-5}$)
O (4.9 $\times 10^{-4}$), P (2.6 $\times 10^{-7}$) and S (1.3 $\times
10^{-5}$). 

Phosphorus abundances have been measured from the mid-infrared \p3
17.9 \mum observations of late stages of stellar remnants such as
planetary nebulae (Pottsch \etal 2008,
Pottasch and Bernard-Salas 2008, Otsuka \etal 2011) 
and from ground based observations
of FGK dwarf stars (Maas \etal 2017).
Whereas phosphorus, with an odd atomic number Z=15, can be synthesized
during the AGB phase of low-mass stars, it
is thought to be mainly 
produced in evolutionary stages of massive stars, before and during
supernovae explosion by neutron capture with silicon. An
analysis 
using singly ionized {\sc P\,ii} lines found up to
100 times the P/Fe ratio in the young core-collapse SNR Cassiopeia A
than the Milky Way average (Koo \etal 2013).  
Phosphorus is chemically very reactive, so its low gas phase abundance may
also be difficult to determine due to dust and grain formation.

 Despite its astrophysical and increasing astrobiological importance, 
theoretical spectral analysis is hampered by
the paucity of radiative and collisional atomic data for phosphorus
ions. It is surprising that very little data for the important
low ionization stages of P-ions in stellar and nebular sources
are available,
relative to nearly all other first and second row elements (viz. Pradhan
and Nahar 2011).
Electron
impact excitation cross sections for {\sc P\,i} have been calculated
in the Born approximation for excitations from the ground $3p$
up to several $n\ell$ sub-orbitals (Ganas 1998). Elaborate Dirac 
R-Matrix calculations for photoionization of low-lying ground and
metastable levels of {\sc 
P\,iii} have been done over a small photon
energy range, with good agreement
with experimental measurements (Wang \etal 2016). 
Recently, sophistcated
Breit-Pauli R-Matrix (BPRM) calculations have been carried out for
photoionization of a large number of {\sc P\,ii} levels
using an 18-level coupled channel
wavefunction expansion for {\sc P\,iii} (Nahar \etal 2017a,b,c);
very good agreement was found with the experimental {\sc P\,ii}
photoionization cross sections  measured at the Berkeley
{\it Advanced Light Source}, particularly for the detailed resonance
structures in the near-threshold region. These earlier works forms the
basis for the calculations reported in this {\it Letter}.

 There are no other previous calculations for collisional excitation of
low ionization stages of P-ions.
We also develop an atomic physics framework for astrophysical
spectral diagnostics
in nebular environments as function
of temperature, density and ionization equilibrium, potentially
leading to more accurate abundance determination. 

\section{Phosphorus abundance analysis}

 Previous works are based on observational analysis of relative
intensities of phosphorus lines compared to other elements. 
For example, nebular
abundances have been determined in planetary nebulae (PNe) 
NGC 3242 and NGC 6369 
from mid-IR observations of \p3 forbidden 17.9 \mum line
using the {\it Infrared Spectrograph} aboard the
{\it Spitzer Space Telescope} and the {\it Short Wavelength
Spectrograph} on the {\it Infrared Space Observatory}
(Pottasch and Bernard-Salas 2008) and NGC 2392 (Pottasch \etal 2008).
A comparison of abundances for NGC
3242 shows phosphorus underabundance in a number of PNe,
by a factor of 5 relative to solar,
and more than a factor of 3 in NGC 6369. The large discrepancies are
attributed to dust formation (Pottasch and Bernard-Salas 2008).
However, the results are model dependent
since they entail atomic parameters and ionization fractions not
known to high precision. 

 For a single observed line the ion abundance may be derived from the
measured intensity ratio under certain conditions.
Relative to recombination line H$\beta$,  we may write
(Pottasch and Beintema 1999)

\be \frac{I_{ion}}{I_{H\beta}} = N_e \frac{N_{ion}}{N_{p+}}
\frac{\lambda_{H\beta}}{\lambda_{ji}} \frac{A_{ji}}{\alpha_R(H\beta)}
\left( \frac{N_j}{N_{ion}} \right) \ee
 
\noindent where N$_{ion}$ is the ionic abundance, N$_j$ is the upper level
population, A$_ji$ is the Einstein decay rate between levels $j \rightarrow
i$, and $\alpha_R$ is the H$\beta$-recombination coefficient.
The present case of \p3 is similar to the
well-known {\sc C\,ii} 157 \mum line emitted via the $2s^22p
^2P^o_{1/2-3/2}$ transition (Blum and Pradhan 1991).
Theoretically,
we write line emissivity for the \p3 FIR transition formed with a
given phosphorus abundance as (Pradhan and Nahar 2011)

$$ \epsilon(17.9 \mu m)  = \frac{h\nu A (^2P^o_{3/2} - ^2P^o_{1/2})}{4\pi}
\times \frac{N(^2P^o_{3/2})}{\sum_i N_i(({\sc P\,III})} 
\times \frac{n({\sc P\,III})}{n(P)}$$
\be \times \frac{n(P)}{n(H)} \times n(H) \ \ ergs/cm^3/sec.  \ee

\noindent The sum in the denominator on the RHS of Eq. (2) refers to all levels
included in the atomic model. Calculating the level populations requires
rate coefficients for contributing atomic processes that may be
due to recombination-cascades, electron impact excitation and
fluorescent excitation from an external radiation field. In addition,
Eq. (2) also depends on ionization balance for existing states in the
plasma, as well as the elemental abundance itself -- the quantity to be
determined. 
However, if the
two closely-spaced levels are effectively de-coupled from other levels 
in the ion then
a simple expression gives the emissivity in terms of only the electron
impact excitation rate coefficient and transition energy $h\nu$,

\be \epsilon (^2P^o_{3/2}-^2P^o_{1/2}) = N_e N_{ion} 
q(^2P^o_{3/2}-^2P^o_{1/2}) h\nu / 4\pi. \ee

\noindent Eq. (3) implicitly assumes that all
 excitations to the upper level
$^2P^o_{3/2}$ would be followed by downward decay to the ground state
 $^2P^o_{1/2}$, leaving the temperature-dependent
electron impact excitation rate coefficient $q$ as the only important
quantity to be
calculated. That, in turn, is related to the Maxwellian averaged
effective collision strength $\Upsilon_{ij}(T_e)$ as

\be q_{ij}(T_e) = \frac{g_j}{g_i} q_{ji} e^{-E_{ij}/kT_e} =
\frac{8.63 \times 10^{-6}}{g_i T^{1/2}} \Upsilon(T_e), \ee

\noindent and

\be
\Upsilon_{ij}(T_e) = \int_0^{\infty} \Omega_{ij} (E)
\exp(-E/kT_e) d(E/kT_e), \ee

\noindent where $E_{ij}$ is the energy difference and $\Omega_{ij}$ is the
collision strength for the transition $i \rightarrow j$. The
exponentially decaying Maxwellian factor implies that at low
temperatures only the very low energy $\Omega_{ij}(E))$ would
determine the $\Upsilon(T_e)$. Furthermore, the detailed $\Omega(E)$ 
is generally
a highly energy-dependent function due to autoionizing
resonances, which leads to temperature sensitivity in
the rate coefficient q($T_e$) via $\Upsilon(T_e)$ as in Eqs. (4-5). 

     \section{Theory and computations}
A brief theoretical description of the calculations is given.
In particular, we describe relatvistic effects and the representation of
the \eion system.

\subsection{Relativistic fine structure} 
The relativistic Hamiltonian (Rydberg units) in the Breit-Pauli R-matrix (BPRM)
approximation is given by

\begin{equation} 
\begin{array}{l}
H_{N+1}^{\rm BP} = \\ \sum_{i=1}\sp{N+1}\left\{-\nabla_i\sp 2 -
\frac{2Z}{r_i}
+ \sum_{j>i}\sp{N+1} \frac{2}{r_{ij}}\right\}+H_{N+1}^{\rm mass} + 
H_{N+1}^{\rm Dar} + H_{N+1}^{\rm so}.
\end{array}
\end{equation}
where the last three terms are one-body relativistic corrections of the
Breit interaction, respectively:
\begin{equation} 
\begin{array}{l}
{\rm the~mass~correction~term},~H^{\rm mass} = 
-{\alpha^2\over 4}\sum_i{p_i^4},\\
{\rm the~Darwin~term},~H^{\rm Dar} = {Z\alpha^2 \over
4}\sum_i{\nabla^2({1
\over r_i})}, \\
{\rm the~spin-orbit~interaction~term},~H^{\rm so}= Z\alpha^2 
\sum_i{1\over r_i^3} {\bf l_i.s_i}.
\end{array} 
\end{equation}

\subsection{Wavefunction representation and calculations}
 Based on the coupled channel approximaton, the R-matrix method (Burke
2011) entails a wavefunction expansion of the \eion
system in terms of the eigenfuctions for the target ion. In the present
case we are interested in low-lying FIR transition within the ground
configuration $3s^23p$ and the next excited configuration $3s3p^2$. 
Therefore we confine ourselves to an accurate wavefunction
representation for the
first 18 levels dominated by the {\it spectroscopic}
 configurations $[1s^2,2s^2,2p^6] 3s^23p (^2P^o_{1/2,3/2)}), 3s3p^2 
(^4P_{1/2,3/2,5/2}, ^2D_{3/2,5/2}, ^2S_{1/2}, \\ ^2P_{1/2,3/2}), 3s^23d
(^2D_{3/2,5/2}), 3s^24s (^2S_{1/2}), 3s^24p (^2P^o_{1/2,3/2}), \\ 3p^3
(^4S^o_{3/2},^2D^o_{3/2,5/2},^2P^o_{1/2,3/2})$. The atomic structure
calculations using the code
SUPERSTRUCTURE (Eissner \etal 1974), and the BPRM calculations
are described in Nahar \etal (2017a). 
The calculated and experimentally
observed energies generally agree to within 5\% for all levels; the
relatively small and sensitive fine structure splitting differs by 15\%
(Nahar 2017a).

 Fig.~1 presents the Grotrian energy level diagram of {\sc P\,iii}. As
noted above, the ground state $3s^23p ^2P^o_{1/2,3/2}$
fine structure is well
separated by about 0.6 Ry or 7 eV from the next excited $3s3p^2$
 configuration terms and levels (Nahar \etal 2017a;
see Pradhan and Nahar (2011) for a general description of
atomic processes and calculations). By comparison, in {\sc C\,ii} it is less
than 0.4 Ry, and approximation in Eq. (4) has been utilized assuming
that the Boltzmann factor $exp(-E_{ij}/kT)$ effectively de-couples the
electron impact excitation of the forbidden FIR transition from higher
levels of the ion (Eq.~4). For example, at $T_e = 10^4$K we have
exp(-E/kT) $\approx$ 
exp(-16E) and the value of 
$q(T_e)$ for allowed UV transitions is orders of
magnitude lower compared to the FIR transition.   
Even though the observed and experimental
values are close, a small difference in resonance positions relative to
threshold can introduce a significant uncertainty in the effective
collision strengths. 
The observed energies were substituted
for theoretical ones in order to reproduce the threshold
energies more accurately. This is of particular importance for
excitation at low temperatures dominated by near-threshold resonances.

\begin{figure} 
\centering  
\includegraphics[width=\columnwidth,keepaspectratio]{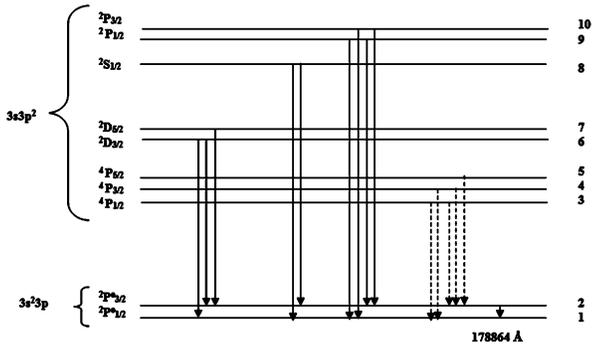}
\caption{Energy diagram of {\sc P\,iii} showing the ground
$3s^23p$ and the first excited configuration $3s3p^2$ levels. 
The energy separation of
the ground state fine structure $^2P^o_{1/2,3/2}$ transition at 17.9
\mum is 0.0051 Ry, and well separated from the dipole allowed
 UV transitions of the $^2P^o_{1/2,3/2} - ^2D, ^2S, ^2P$ multiplets
between  $915-1345 \AA$ with E $>$ 0.68 Ry.
\label{energies}} 
\end{figure}

 The BPRM collision strengths were computed 
up to 5 times the energy of the highest level in the
atomic calculations, $3p^3 (^4S^o_{3/2})$ at 1.45 Ry. Particular care
is taken to
test and ensure convergence of collision strengths with respect to
partial waves and energy resolution. Total \eion symmetries up to
(LS)J$\pi$ with J $\leq 19.5$ were included in the calculations, though
it was found that the collision strengths for forbidden
transitions converged for J $\leq 7$. An energy mesh of $\Delta E
\sim 10^{-5}$ Rydbergs was used to resolve the near-thresold resonances. The
resonances were delineated in detail prior to averaging over the
Maxwellian distribution. 

\section{Results}

 We describe the two main sets of results for the FIR and the 
UV transitions, as well as the diagnostic lines. Collision strengths
have been computed for all 153 transitions among the 18 
{\sc P\,iii} levels. Selected results are presented below;
no previous data are available for comparison. 

\subsection{The forbidden 17.9 \mum transition}

 The calculated fine structure collision strength is shown in Fig. 2a,
that exhibits considerable autoionizing resonance structures and
energy dependence
throughout the range up to the highest level of the {\sc P\,iii} ion
included in the BPRM wavefunction expansion, $E(3p^3 \ ^2P^o_{3/2})$ = 1.45 Ry.
The fall-off for E $>$ 1.0 
Ry indicates that the collision strength is much lower at higher
energies, and has converged for this forbidden transtion. 

One particularly noteworthy feature is that the 17.9 \mum FIR
transition is very strong, with large collision
strengths and resonances just above the
excitation threshold at E $\approx$ 0.1 Ry. That yields a
maximum effective collision strength
$\Upsilon(T_e) > 2.0$ between $10^3-10^4$K; 
by comparison the strong 157 micron transition in {\sc
C\,II} has a value of $\sim$1.6 (Blum and Pradhan 1991). Consequently, the
exciation rate coefficient and emissivity (Eqs. 3, 4)
would indicate strong observable intensity relative to other FIR
lines from other elements (viz. Pottasch and
Bernard-Salas). 
In addition, the energy dependence of $\Omega(E)$ in Fig.~3(a) leads to
variation of more than a factor of 3 in $\Upsilon(T_e)$ in
Fig.~3b. Therefore, the intensity of the line is a sensitive indicator of
temperature in the typical nebular range of $10^2
- 10^5$K, encompassing spectral formation in important sources such
as PNe and SNRs.

\begin{figure} 
\centering  
\includegraphics[width=\columnwidth,keepaspectratio]{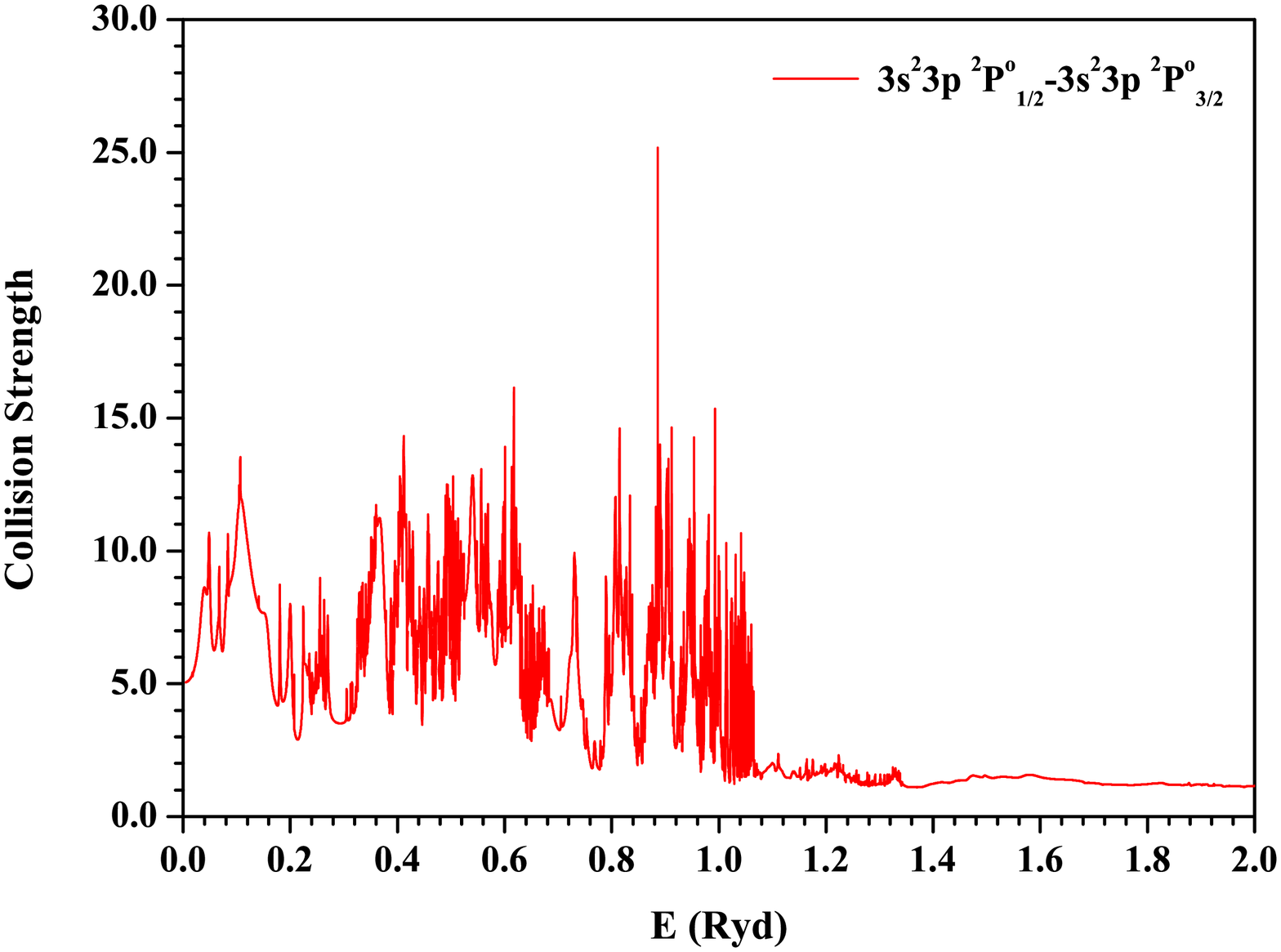}
\includegraphics[width=\columnwidth,keepaspectratio]{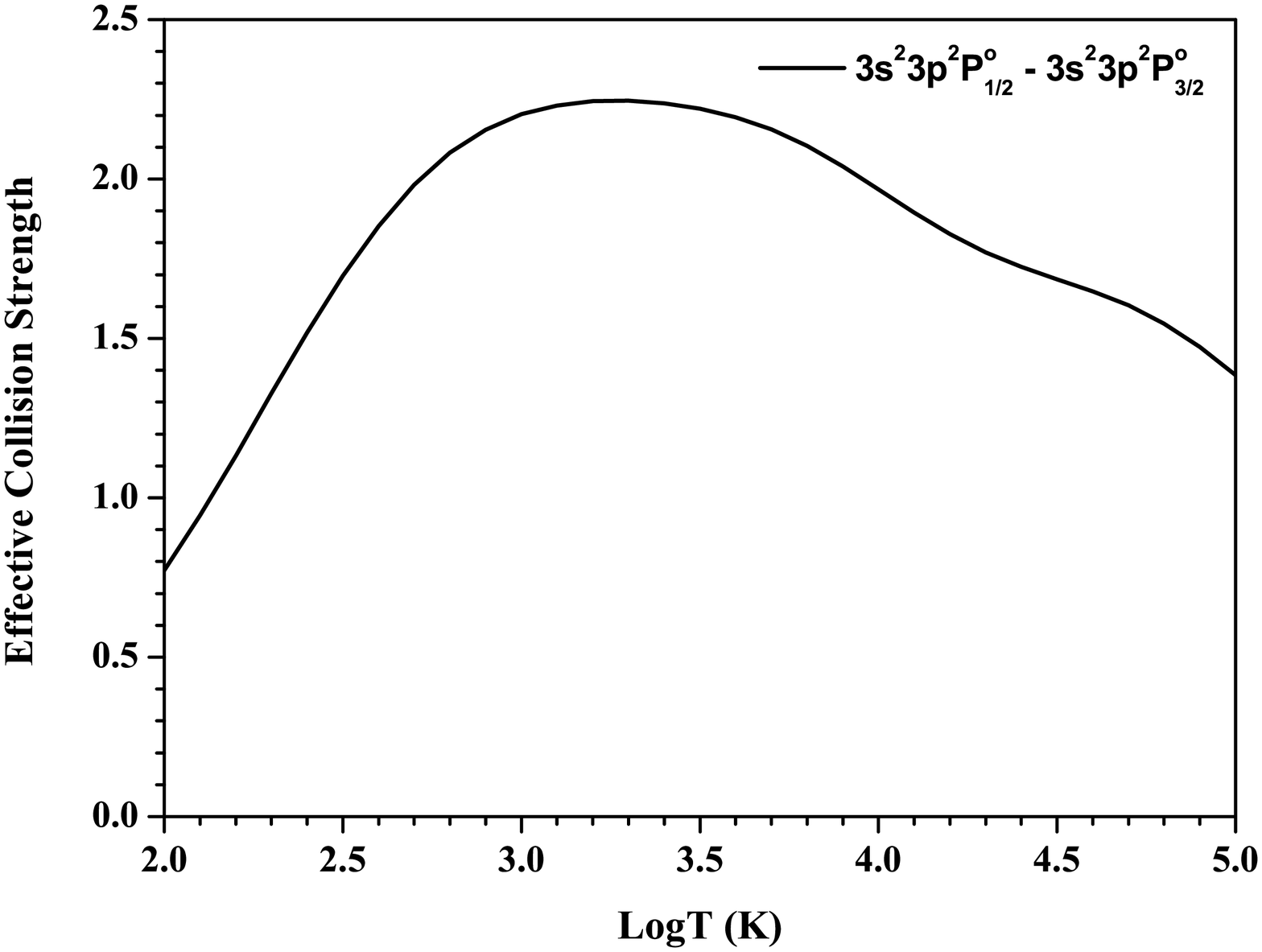}
\caption{(a) Collision strength for the 17.9 \mum \p3 IR fine structure
transition. High resolution at near-threshold
energies is necessary for accuracy in rate coefficients at low
temperatures. (b) Maxwellian averaged effective collision strengths
$\Upsilon(T_e)$ (Eq. 5).
There is a factor of 3 or more variation 
broadly peaking at typical nebular temperatures Te$_e > 10^3$K.
structures. \label{fir}} 
\end{figure}

\subsection{Allowed UV transitions}

 There are a number of intercombination and 
dipole allowed E1 transitions between the
odd parity ground state fine structure levels $3s^23p  (^2P^o_{1/2-3/2})$
and the even parity $3s3p^2 (^4P_{1/2,3/2,5/2},^2D_{3/2,5/2}, ^2S_{1/2},
^2P_{1/2,3/2})$ levels. However, laboratory and theoretical radiative data 
for measured wavelengths and Einstein A-values available from the
National Institute of Standards and Technology show only the three
transitions in the $^2P^o_{1/2,3/2} - ^2D_{3/2,5/2}$.
Fig.~3a presents sample collision strengths for fine struture components
of dipole transitions in the three allowed multiplets.
The BPRM calculations again
show resonance 
structures below the highest target ion threshold at 1.45 Ry due to
low-$\ell$ partial waves included in the calculations with $\ell_o \approx
10$. As these
are E1 transitions, the collision strengths rise with increasing energy
owing to divergent higher partial wave contributions $\ell > \ell_o$. 
The general form in the high energy region may be approximated by the
Bethe formula $\Omega \sim a lnE$, where $a$ is related to the dipole
oscillator strength, assuming high-$\ell$ collisions as radiatively
induced (Pradhan and Nahar 2011). We therefore, match the BPRM collision
strengths at 1.45 Ry to the Bethe expression. While there may be some
uncertainty in the vicinity of this energy region, the overall behaviour
of the collision strengths in Fig.~3a appears to be correct (c.f.
Blum and Pradhan (1991) for C~II collision strengths for similar
transitions). The effective collision strengths $\Upsilon(T_e)$ in
Fig.~3b show the expected rising behaviour with temperature, typical of
allowed transitions.

\begin{figure} 
\centering
\includegraphics[width=\columnwidth,keepaspectratio]{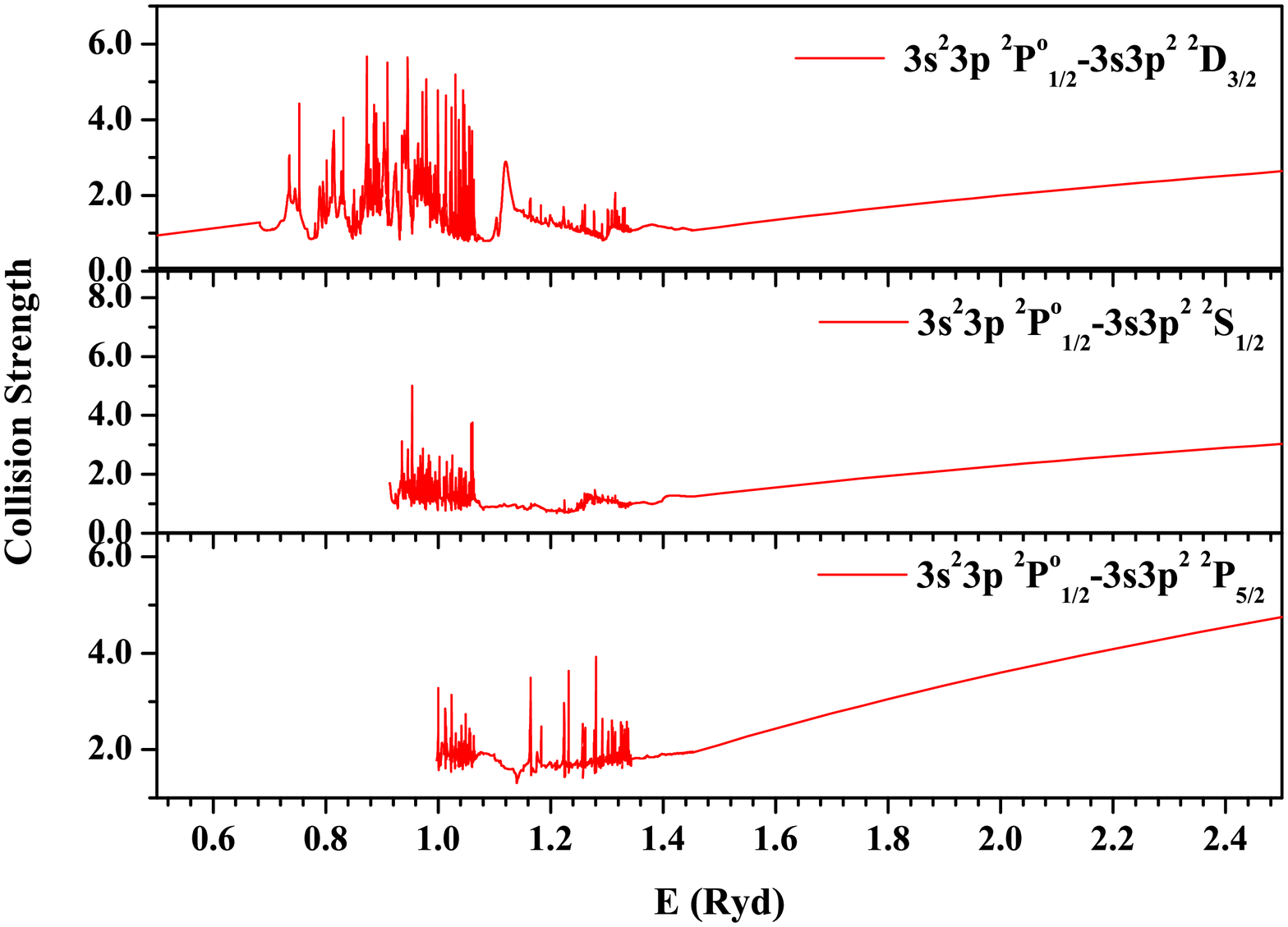}
\includegraphics[width=\columnwidth,keepaspectratio]{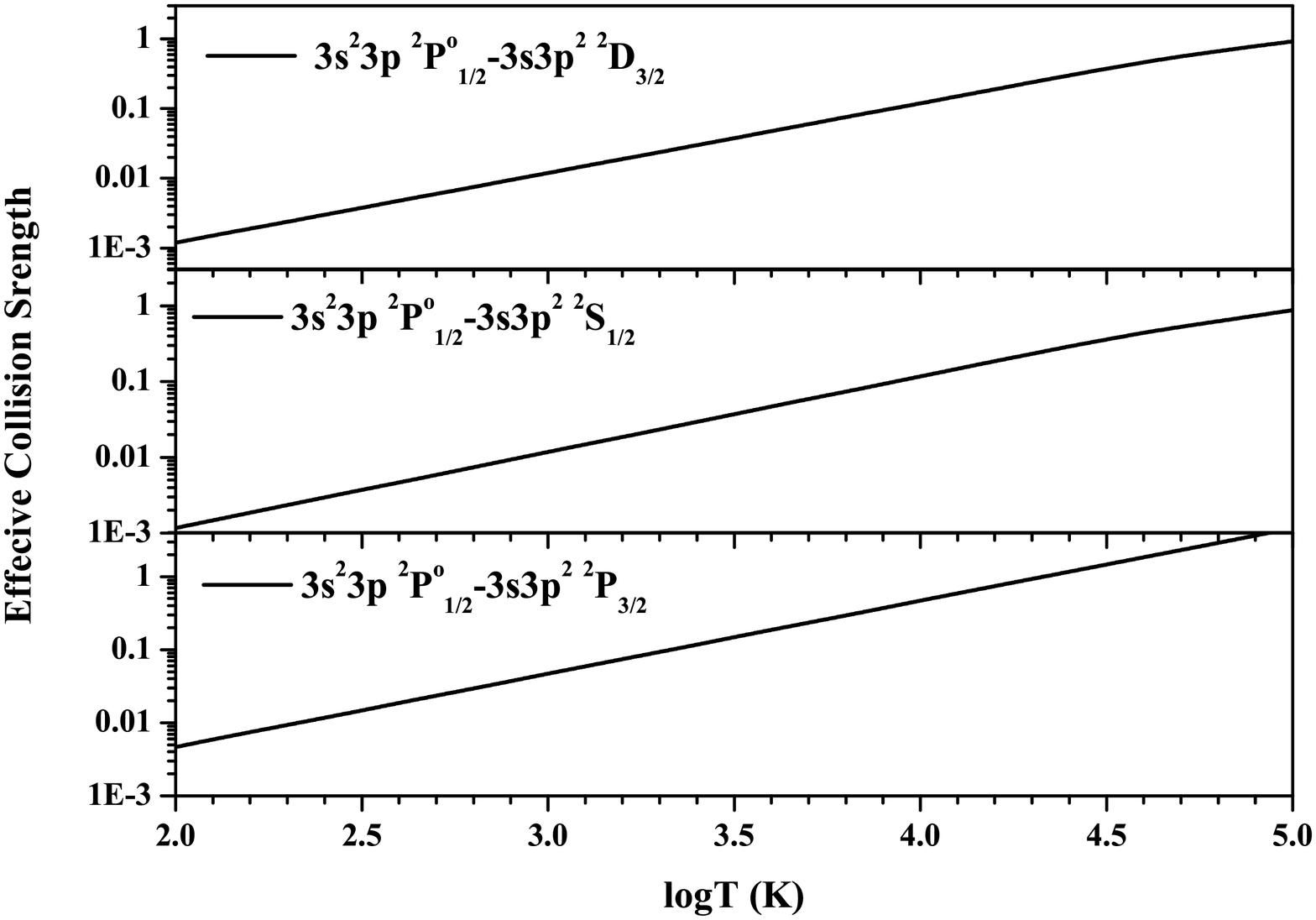}
\caption{(a) Collision strengths $\Omega(E)$ for sample UV fine structure
transitions from the ground level $^2P^o_{1/2} \rightarrow ^2D_{3/2},
^2S_{1/2}, ^2P_{3/2}$ 
For E $>$ 1.5 Ry the Coulomb-Bethe form $\Omega(E) \sim ln E$ is employed, 
typical of
dipole allowed transitions at high energies and partial waves. (b)
Maxwellian averaged effective collision strengths $\Upsilon(T_e)$
for the transitions transitions in (a).
\label{uv}}
\end{figure}

\subsection{Maxwellian averaged collision strengths}

 In Table~1 we present the effective collision strengths (Eq. 3) for the
four FIR and UV transitions reported herein.
The tabulation is carried out at a range of temperatures typical of
nebular environments $10^2-10^5$K. It is striking how much stronger the
forbidden FIR 17.9 \mum is relative to the allowed UV transitions, and
dominates collisional excitation to all other levels by up to two orders of
magnitude for temperatures between 100-10,000 K, although the values
become comparable towards higher temperatures as shown in Fig.~3b. That
further numerically supports the approximation that the FIR line intensity
may be little affected by excitation to higher levels.

\begin{table*}
\begin{minipage}{188mm}
\caption{Effective Maxwellian averaged collision strengths for FIR and
UV transitons in {\sc P\,iii}}
\begin{tabular}{cccccccccc}
\hline
$LogT$(K) & $^2P^o$ & $^2P^o-^2D$ & $^2P^o-^2S$ &$^2P^o-^2P$ & 
$LogT$(K) & $^2P^o$ & $^2P^o-^2D$ & $^2P^o-^2S$ &$^2P^o-^2P$\\
 J-J' & 1/2-3/2 & 1/2-3/2 & 1/2-1/2 & 1/2-3/2 & J-J' & 1/2-3/2 &
1/2-3/2 & 1/2-1/2 & 1/2-3/2\\
$\lambda$ &  17.9 $ \mu$m &   1334.8 $\AA$ & 998.6 $\AA$ & 914.5 $\AA$
& $\lambda$ &  17.9 $ \mu$m &   1334.8 $\AA$ & 998.6 $\AA$ & 914.5 $\AA$ \\
\hline
 
	2.0&	7.73(-1)&	1.20(-3)&	1.18(-3)& 	1.13(-3)& 
	3.6&	2.19&		4.76(-2)&	4.68(-2)&	4.51(-2)\\
	2.1&	9.45(-1)&	1.50(-3)&	1.48(-3)&	1.43(-3)&
        3.7&	2.16& 		5.99(-2)&	5.89(-2)&	5.67(-2)\\
	2.2&	1.13&   	1.90(-3)&	1.86(-3)&	1.79(-3)&
        3.8&	2.10& 		7.54(-2)&	7.41(-2)&	7.14(-2)\\
	2.3&	1.33&		2.39(-3)&	2.34(-3)&	2.26(-3)&
        3.9&	2.04& 		9.50(-2)&	9.33(-2)&	8.99(-2)\\
	2.4&	1.52&    	3.00(-3)&	2.95(-3)&	2.84(-3)&
        4.0&	1.97& 		1.20(-1)&	1.18(-1)&	1.13(-1)\\        
	2.5&	1.70& 	        3.78(-3)&	3.72(-3)&	3.58(-3)&
        4.1&	1.90& 		1.51(-1)&	1.48(-1)&	1.43(-1)\\
	2.6&	1.85& 		4.76(-3)&	4.68(-3)&	4.51(-3)&
        4.2&	1.83& 		1.90(-1)&	1.86(-1)&	1.79(-1)\\
	2.7&	1.98& 		5.99(-3)&	5.89(-3)&	5.67(-3)&
	4.3&	1.77& 		2.39(-1)&	2.34(-1)&	2.26(-1)\\
	2.8&	2.08& 		7.54(-3)&	7.41(-3)&	7.14(-3)&
	4.4&	1.73& 		3.00(-1)&	2.93(-1)&	2.84(-1)\\
	2.9&	2.16& 		9.50(-3)&	9.33(-3)&	8.99(-3)&
        4.5&	1.69& 		3.75(-1)&	3.63(-1)&	3.56(-1)\\
	3.0&	2.20& 		1.20(-2)&	1.18(-2)&	1.13(-2)&
	4.6&	1.65& 		4.63(-1)&	4.43(-1)&	4.45(-1) \\
	3.1&	2.23& 		1.51(-2)&	1.48(-2)&	1.43(-2)&
	4.7&	1.60& 		5.62(-1)&	5.32(-1)&	5.55(-1) \\
	3.2&	2.24& 		1.90(-2)&	1.86(-2)&	1.79(-2)&
  	4.8&	1.55& 		6.70(-1)&	6.32(-1)&	6.91(-1) \\
	3.3&	2.25& 		2.39(-2)&	2.34(-2)&	2.26(-2)&
	4.9&	1.47& 		7.87(-1)&	7.45(-1)    &	8.64(-1) \\
	3.4&	2.24& 		3.00(-2)&	2.95(-2)&	2.84(-2)&
        5.0&	1.39& 		9.19(-1)&	8.82(-1)    &	1.09 \\
	3.5&	2.22& 		3.78(-2)&	3.72(-2)&	3.58(-2)&&&&\\
	
\hline
\end{tabular}
\end{minipage}
\end{table*}

\subsection{Discussion}
 The results reported herein should enable spectral diagnostics of both
the \p3 forbidden 17.9 \mum line as well as UV transitions with a
practically complete 18-level collisional-radiative atomic model. 
The FIR and UV lines can
not be observed with the same spectrometer and their spectral formation
may be governed by different physical conditions, as well as subject to
extinction
that is highly wavelength dependent and would differntially attenuate
line intensities. Some temperature dependence may be
deduced from the energy behaviour inherent in the
collision strengths data presented, and derived line emissivities.
Based on extensive benchmarking of R-matrix data with
experiments, we estimate the accuracy of the effective collision strengths 
between 10-20\%.

We may calculate nebular phosphorus abundance as outlined in Section 2
Eqs. 1-5, based on the
\p3 17.9 \mum line intensity ratio relative to H$\beta$.
We assume a temperature $10^4$K, density
$10^4 \ cm^{-3}$, transtion energy $h\nu$ = 0.069 eV,
solar P-abundance, and ionic ratio P~III/P = 0.33. Using 
$\Upsilon(10^4K)$ from Table 1, the
rate coefficient $q = 7.77 \times 10^{-8} \ cm^{3}/sec$ 
and 
$(4 \pi / N_p N_e) \epsilon(17.9 \mu m) = 7.33 \times 10^{-28} \
ergs/cm^3/sec$. Nebular recombination H$\beta$ line
intensities $(4\pi/N_pN_e) j(H\beta)$ 
are: $8.3 \times 10^{-26}$ (Case A) and
$1.24 \times 10^{-25} \ ergs \ cm^3 \sec$ (Case B).
Therefore, $\epsilon(17.9\mu m)/H\beta \ = 8.8 \times
10^{-3}$ (Case A) and $5.9 \times 10^{-3}$ (Case B) respectively.  
These $\epsilon(17.9\mu m)/H\beta$ line ratios
lie in the range observed in several PNe,
but the P-abundances heretofore derived
are factor of 3-4 lower than solar (viz. Pottasch and Bernard-Salas
2008); present work may yield higher abundances. 

Further refinements can be
made by considering additional atomic processes such as
level-specific (e~+~P~IV) $\rightarrow$ {\sc P\,iii}
recombination-cascades, and flourescent excitation from an external
radiation field such as in PNe central stars with $T_{rad} \approx
80,000-120,000$K. A more elaborate calculation can be done
using Eq.~(2) that would combine the collisional-radiative model with a
photoionization model that describes P-ionization states more accurately
than, say, the {\sc P\,iii}/P value of 0.33 assumed above. 
However, these improvement would require extensive new atomic
calculations
for photoionization and \eion recombination (e.g. Nahar \etal 2017).
 
 An interesting possibility is that
of laser action in the 17.9 \mum line, similar to that explored for the
{\sc C\,ii} 157 micron transition (Peng and Pradhan 1994). 
Population inversion may occur 
owing to the extremely small magnetic dipole (M1) radiative decay rate
A($^2P^o_{3/2}-^2P^o_{1/2}$) = 1.57$\times 10^{-3}/sec$ (NIST {\it Atomic
Spectral Database}: www.nist.gov). 
Equating \ne $q$ = A, we obtain \ne = 2.0 $\times 10^4$
cm$^{-3}$. Therefore, at electron
densities \ne $> 10^4$ cm$^{-3}$, electron impact excitation
exceeds spontaneous decay, and
population inversion and laser emission may 
occur in higher density SNRs or other sources.

\subsection{Conclusion}

 Accurate collision strengths including fine structure with relativistic
effects are computed for diagnostics of
the \p3 forbidden FIR and allowed UV lines
to enable a more precise re-examination of
phosphorus abundance.
The results show signficant temperature dependence that should
provide additional information on the physical environment and spectral
formation. In particular, this work suggests searches for the \p3 FIR
line using {\it Spitzer} IRS data and abundance determination.
Further work is in progress on photoionization and
collisional excitation of P-ions relevant to this investigation.
 All data are available from the authors and archived in 
 S. N. Nahar's database NORAD at:
www.astronomy.ohio-state.edu$\sim$nahar, and TIPTOPBase at
the Opacity Project/Iron Project webpage:
http://cdsweb.u-strasbg.fr/OP.htx. 

\section*{Acknowledgments}
 The computational work was 
carried out at the Ohio Supercomputer Center in Columbus 
Ohio. This work was partially supported by the Astronomy
Division of the U.S. National
Science Foundation (SNN and AKP), and from
the Indo-US Science and Technology Forum and Science and Engineering
Research Board, Government of India (RN).

\label{lastpage}

\frenchspacing
\def\aa{{\it Astron. Astrophys.}\ }
\def\aasup{{\it Astron. Astrophys. Suppl. Ser.}\ }
\def\adndt{{\it Atom. data and Nucl. Data Tables.}\ }
\def\aj{{\it Astron. J.}\ }
\def\apj{{\it Astrophys. J.}\ }
\def\apjs{{\it Astrophys. J. Supp. Ser.}\ }
\def\apjl{{\it Astrophys. J. Lett.}\ }
\def\baas{{\it Bull. Amer. Astron. Soc.}\ }
\def\cpc{{\it Comput. Phys. Commun.}\ }
\def\jpb{{\it J. Phys. B}\ } 
\def\jqsrt{{\it J. Quant. Spectrosc. Radiat. Transfer}\ }
\def\mn{{\it Mon. Not. R. astr. Soc.}\ }
\def\pasp{{\it Pub. Astron. Soc. Pacific}\ }
\def\pra{{\it Phys. Rev. A}\ }
\def\pr{{\it Phys.  Rev.}\ } 
\def\prl{{\it Phys. Rev. Lett.}\ }

\bibliography{ms}
\end{document}